\documentclass[sigconf]{acmart}

\AtBeginDocument{%
  }

\setcopyright{acmlicensed}
\copyrightyear{2018}
\acmYear{2018}
\acmDOI{XXXXXXX.XXXXXXX}
\acmConference[Conference acronym 'XX]{Make sure to enter the correct
  conference title from your rights confirmation email}{June 03--05,
  2018}{Woodstock, NY}
\acmISBN{978-1-4503-XXXX-X/2018/06}

\usepackage{enumitem}
\usepackage{graphicx}
\usepackage{subcaption}
\usepackage{multirow}
\usepackage{array}

\begin{document}

\title{IAT: Instance-As-Token Compression for Historical User Sequence Modeling in Industrial Recommender Systems}


\author{Xinchun Li$^*$, Ning Zhang$^*$, Qianqian Yang$^*$, Fei Teng$^*$, Wenlin Zhao$^*$, Huizhi Yang$^*$ \\ Heng Shi, Linlan Chen, Yixin Wu, Zhen Wang, Daiye Hou, Fei Qin, Lele Yu, Yaocheng Tan$^\dagger$}
\thanks{$^*$ Co-first authors, equal contributions. \\ $^\dagger$ Corresponding author.}
\affiliation{
  \institution{ByteDance}
  \country{}
}
\email{{lixinchun.bu, zhangning.1230, yangqianqian.0920, taolangfei, zhaowenlin, yanghuizhi}@bytedance.com}
\email{{shiheng.sai, chenlinlan, wuyixin.yx, wangzhen.raynar, houdaiye, qinfei.ailab, yulele, tanyaocheng}@bytedance.com}

\renewcommand{\shortauthors}{Li et al.}

\begin{abstract}
  Although sophisticated sequence modeling paradigms have achieved remarkable success in recommender systems, the information capacity of hand-crafted sequential features constrains the performance upper bound. To better enhance user experience by encoding historical interaction patterns, this paper presents a novel two-stage sequence modeling framework termed \textbf{Instance-As-Token} (\textbf{IAT}). The first stage of IAT compresses all features of each historical interaction instance into a unified instance embedding, which encodes the interaction characteristics in a compact yet informative token. Both temporal-order and user-order compression schemes are proposed, with the latter better aligning with the demands of downstream sequence modeling. The second stage involves the downstream task fetching fixed-length compressed instance tokens via timestamps and adopting standard sequence modeling approaches to learn long-range preferences patterns. Extensive experiments demonstrate that IAT significantly outperforms state-of-the-art methods and exhibits superior in-domain and cross-domain transferability. IAT has been successfully deployed in real-world industrial recommender systems, including e-commerce advertising, shopping mall marketing, and live-streaming e-commerce, delivering substantial improvements in key business metrics.
\end{abstract}




\maketitle

\section{Introduction}
\begin{figure}[tb]
  \centering
  \includegraphics[width=\linewidth]{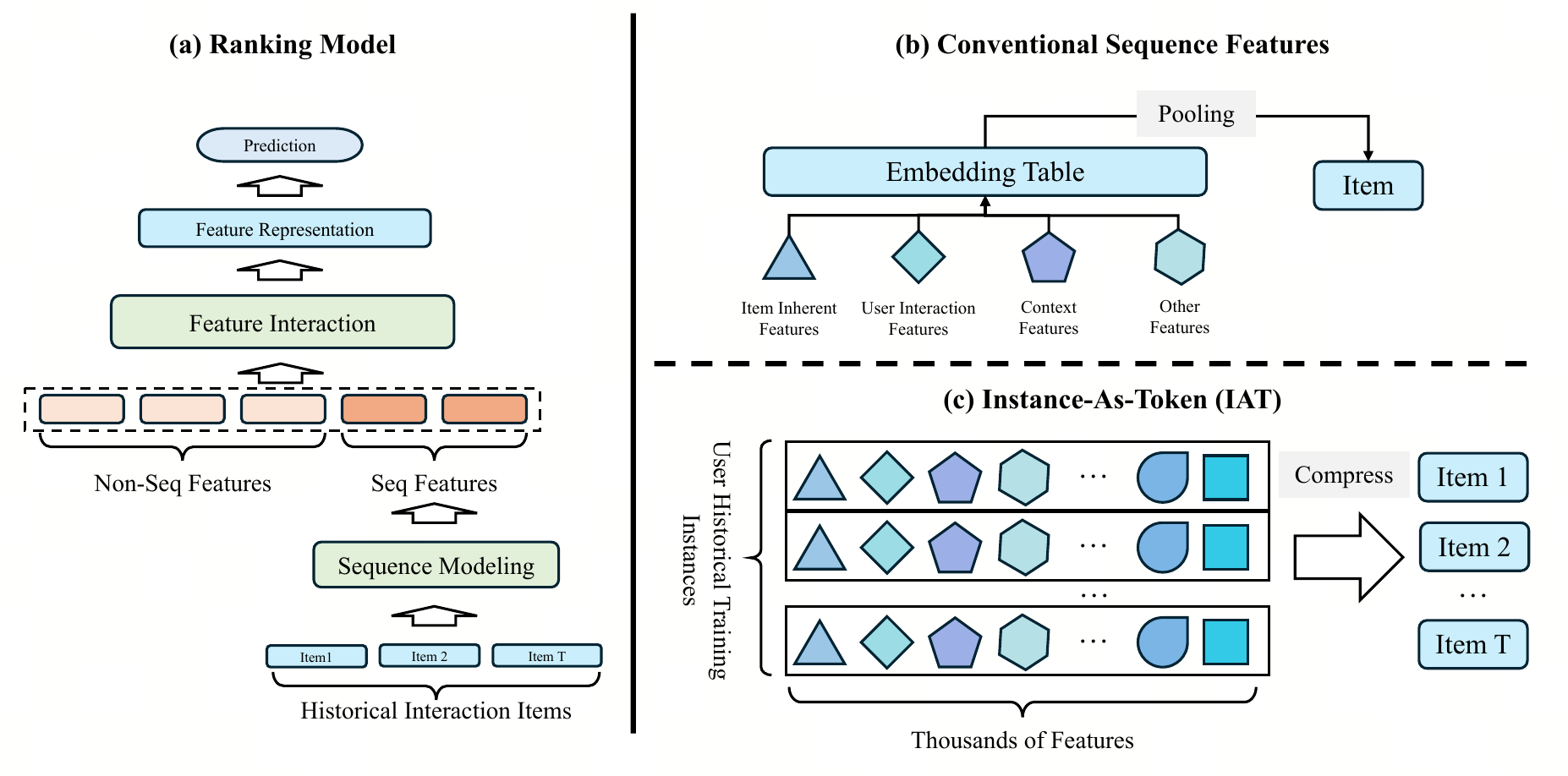}
  \caption{The motivation of IAT. Hand-crafted sequence features limit the further scaling of advanced ranking architectures. Instead, IAT proposes compressing all features within training instances that represent users' historical interactive behaviors into unified tokens for sequence modeling.}
  \label{fig:teaser}
\end{figure}

Modeling short-term and long-term interaction sequences has been widely shown to improve recommendation effectiveness and user satisfaction in modern recommender systems~\cite{DIN,TWINV1,UniSRec,LONGER}. The left side of Fig.~\ref{fig:teaser} shows the mainstream architecture of ranking models. Existing studies focus on designing more sophisticated modeling paradigms~\cite{Wukong,RankMixer,TokenMixerLarge,STCA}, constructing efficient architectures for ultra-long sequences~\cite{TWINV2,TransActV2,ENCODE,LREA,VQL,DMQN,HiSAC}, proposing enhanced feature interaction mechanisms between sequential and non-sequential features~\cite{HiFormer,InterFormer,OneTrans,HyFormer}, and validating scaling laws~\cite{CLIMBER,Scaling1B,SUAN} as well as generative retrieval~\cite{GR,TIGER,OneRec} in recommender systems.

However, these efforts focus primarily on high-level model design and \textit{still rely on low-level sequential feature engineering approaches}. Existing sequential features are typically derived from a set of hand-crafted features as shown in Fig.~\ref{fig:teaser} (b), including inherent item features (e.g., price, category), user-item interaction features (e.g., interaction type, frequency), and context features (e.g., timestamp)~\cite{TWINV1,DV365,TransActV2,UBMSurvey}. On the one hand, resource constraints in storage, transmission, and computation make it difficult to construct such hand-crafted features on a large scale~\cite{Rabbitail,SCRec} and fine-grained features are excluded, resulting in sparse feature representations and degraded modeling performance. On the other hand, sequential feature engineering—including feature design, development, and validation—usually involves a long cycle, leading to high industrial costs for feature expansion~\cite{FeatureTutorial,DIF,Trans2D}. These two aspects limit the information density in historical interaction sequences and further \textit{hinder the effective scaling of modern architectures}~\cite{FDSA,LLaTTE}.


To address the above issues and better facilitate the learning process of upper interaction modules, we propose a novel sequence modeling framework that focuses on \textit{enhancing the information capacity of sequences}. The motivation is to utilize a dense yet informative embedding to represent a specific historical interaction, rather than relying on hand-crafted features that carry sparse and limited information, which is illustrated in Fig.~\ref{fig:teaser} (c). The training instances in modern industrial recommender systems could thoroughly describe historical interaction patterns, commonly containing thousands of features~\cite{ERASE,GR}. We aim to \textit{compress the historical training instances of a user into dense representations and then use these representations as tokens for downstream sequence modeling}. The framework is named \textbf{Instance-As-Token} (\textbf{IAT}), which consists of two stages as follows.

\begin{itemize}[leftmargin=*]
    \item {\textbf{IAT Compression}}. This stage produces compact instance embedding (\textbf{InsEmb}) for each training instance by specific compression mechanisms. As an intuitive solution, we train a \textit{temporal-order source model} by applying compression and decompression layers to the final feature representation of a base model, with the intermediate compressed representations stored as InsEmb. Aside from this, we further propose the \textit{user-order source model} that organizes the training instances in user order and introduces a \textbf{Source Instance Transformer} (\textbf{SIT}) module to accomplish the sequence modeling process. SIT significantly enhances the performance of the source model and simultaneously endows the generated InsEmb with sequence modeling capability, thus demonstrating remarkable performance transferability when applied to downstream models. Several complementary features that describe historical training instances (e.g., multi-task labels) are \textit{optionally} stored along with the generated InsEmb.
    \item {\textbf{IAT Sequence Modeling}}. This stage retrieves the stored informative InsEmb and the optional key features, which are aggregated as Instance Tokens (\textbf{InsToken}) in the downstream models. To avoid future information leakage, the retrieved InsTokens are strictly truncated by the request timestamp. Then, modern architectures such as LONGER~\cite{LONGER} or Transformer~\cite{Transformer} could be adopted for IAT sequence modeling. Benefiting from the informative InsTokens, the downstream model achieves significant performance improvements in both offline evaluations and online A/B tests.
\end{itemize}


In summary, our main contributions are fourfold: \textit{(a) Novel Sequence Feature Engineering}. The proposed IAT presents a novel sequence feature engineering mechanism to replace inefficient and less effective hand-crafted features. \textit{(b) Distinct and Appropriate IAT Compression Schemes}. We propose both temporal-order and user-order compression schemes for generating InsEmb, with the latter aligning well with the downstream model. \textit{(c) Complete and Detailed IAT Procedures}. We elaborate on the procedures of IAT from the perspectives of model architecture, streaming training pipeline, and storage deployment process. \textit{(d) Remarkable Performance Improvements}. Extensive offline experiments validate the advantages of the proposed IAT framework, and real-world industrial recommender systems across multiple scenarios achieve significant online metric gains after the introduction of IAT.

\section{Related Works}
\subsection{Sequence Modeling Paradigms}
Behavior sequence modeling enables capture of dynamic preferences in recommender systems~\cite{DIN,TWINV1,DV365}. Existing studies mainly focus on several key directions to improve the performance, efficiency, and scalability of sequence modeling. Works in~\cite{Wukong,RankMixer,TokenMixerLarge} enhance heterogeneous feature interaction in large-scale ranking models, while those in~\cite{OneTrans,HyFormer} enable sufficient information integration between sequential and non-sequential features. Researches in~\cite{TWINV2,VQL,DMQN,HiSAC} aim to handle longer user sequences via efficient architecture design, such as clustering and sparse attention techniques~\cite{NSA}. Exploring scaling laws~\cite{ScalingLaw,Scaling1B,CLIMBER} in recommender systems and advancing the development of the generative recommendation paradigm~\cite{GR,OneRec,MTGR,OneMall} are also popular research directions. However, these studies \textit{fail to consider the limited information capacity of traditional hand-crafted sequential features}.

\subsection{Sequence Feature Engineering}
The information capacity covered by low-level sequential item features dictates the upper bound of high-level recommendation model performance. Early sequential recommendation works only utilize ID features~\cite{SASRec}, while subsequent studies point out that other sequential features are beneficial for capturing fine-grained preferences~\cite{FDSA}. Some studies further propose advanced methods to fuse side information of items beyond IDs~\cite{DIF,Trans2D,ASIF}. Leveraging multimodal signals of items can effectively improve the generalization ability of recommender systems, including multimodal embedding~\cite{MUSE,LEMUR} and semantic IDs~\cite{OneRec,FORGE}. However, these methods either \textit{remain dependent on hand-crafted features} or \textit{introduce prohibitive overhead} and \textit{suffer from information misalignment}. In contrast, the proposed IAT retains critical information across all features while achieving better alignment with downstream tasks.


\subsection{Two-Stage Ranking Frameworks}
Under the constraints of resource usage and online latency, the two-stage ranking framework~\footnote{The two-stage paradigm does not refer to the conventional retrieval-ranking pipeline, but a two-stage modeling approach employed within the ranking stage.} can \textit{achieve more flexible scaling in model deployment}~\cite{MARM,PretrainUE}. Recent works~\cite{LargeFM,SUAN} propose building large teacher models as foundation models and utilizing their distillation signals to enhance the performance of small-capacity student models. HLLM~\cite{HLLM} employs a two-tier LLM model, where the first item LLM extracts rich content features of items and the second user LLM accomplishes sequence modeling over these item features. LLaTTE~\cite{LLaTTE} points out that semantic features are a prerequisite for scaling and introduces a two-stage architecture that includes an upstream user model generating and caching compressed user embeddings. Our proposed IAT also adopts a two-stage training approach, where the first stage generates dense yet informative instance embeddings and the second stage utilizes these embeddings to achieve better sequence modeling.

\begin{figure}[tb]
  \centering
  \includegraphics[width=\linewidth]{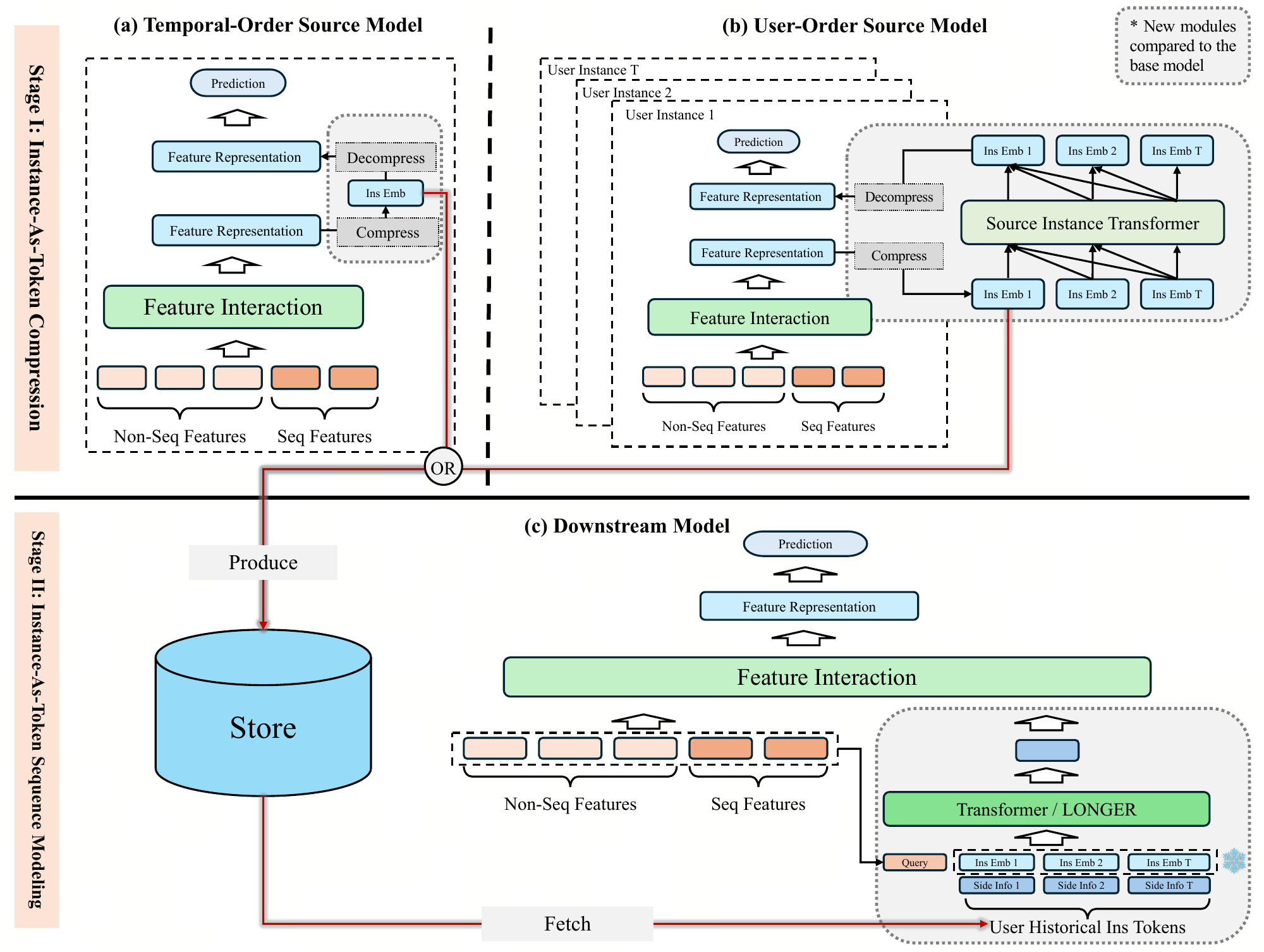}
  \caption{The overall two-stage framework of IAT. The IAT Compression stage optionally builds the designed temporal-order or user-order source model, which generates and stores compact and informative InsEmb. The second stage trains a downstream model that retrieves these InsEmb for sequence modeling.}
  \label{fig:framework}
\end{figure}

\section{Methodology}
\subsection{Problem Statement}
\label{sec:base}
We first introduce the basic architecture of a popular ranking model as the base model. We use lowercase letters to denote data for a single instance, while uppercase letters denote data for a batch. The features commonly consist of sequential and non-sequential features, denoted as $x_{\text{seq}}$ and $x_{\text{non-seq}}$, respectively.
The sequence features usually cover $T$ historical interaction behaviors, i.e., $x_{\text{seq}}=\{s_1,s_2,\cdots,s_T\}$, which are first processed by a sequence modeling architecture $\mathcal{F}_{\text{seq}}$ and then combined with other non-sequential features for deeper feature interaction, i.e.,
\begin{equation}
    h = \mathcal{F}_{\text{interaction}}\left(x_{\text{non-seq}}, \mathcal{F}_{\text{seq}}(x_{\text{seq}})\right), \label{eq:fi}
\end{equation}
where $h$ denotes the compact feature representation which is subsequently used to predict task objectives (e.g., click-through rate (CTR) or conversion rate (CVR)) as follows:
\begin{equation}
\hat{y} = P(y=1 \mid u, v, h)
\end{equation}
where $y \in \{0,1\}$ denotes whether the user $u$ will be interested in the item $v$. The model is trained with the binary cross-entropy (BCE) loss:
\begin{equation}
\mathcal{L} = -\frac{1}{|\mathcal{D}|} \sum_{(u,v,x,y) \in \mathcal{D}} \left(y\log\hat{y} + (1-y)\log(1-\hat{y})\right),
\end{equation}
where $\mathcal{D}$ denotes the set of training instances, and $x=(x_{\text{seq}},x_{\text{non-seq}})$. 

\subsection{Overall Two-Stage IAT Framework}
Existing studies focus on improving the scaling capacity of $\mathcal{F}_{\text{seq}}$ or $\mathcal{F}_{\text{interaction}}$, while \textit{ignoring the information bottleneck caused by the hand-crafted $x_{\text{seq}}$}. As illustrated in Fig.~\ref{fig:teaser} (c), a user's training instance completely describes historical interactions through thousands of features, which motivates the proposed Instance-As-Token (IAT) — a novel sequence modeling framework shown in Fig.~\ref{fig:framework}. IAT consists of two stages: (1) \textbf{Stage I (IAT Compression)}. We apply two types of compression modules to the base model, optionally obtaining the \textit{temporal-order or user-order source model}. The source models process training instances and generate compact instance embeddings (\textbf{InsEmb}), which are stored in a centralized repository. This stage compresses rich-feature instances into low-dimensional embeddings, enabling efficient storage and transmission of comprehensive instance information. (2) \textbf{Stage II (IAT Sequence Modeling)}. The \textit{downstream model} constructs instance tokens (\textbf{InsToken}) by combining the InsEmb sequence with complementary key side information for sequential modeling. The following section details how to construct source models and downstream models based on the ranking model introduced in Sect.~\ref{sec:base}.



\subsection{Stage I: IAT Compression}
This stage constructs \textit{source models} to obtain compressed instance representations. We focus on the key component for InsEmb generation, namely two distinct compression schemes.
\subsubsection{Temporal-Order Source Model}
To generate a compact representation of an instance, an intuitive solution is to simply utilize the representation obtained by complex feature interaction, i.e., the $h$ defined in Eq.~\ref{eq:fi}. However, $h$ typically denotes a high-dimensional feature representation with thousands of dimensions. Directly storing such a representation inevitably introduces prohibitive storage and transmission overhead. Hence, we add a compression and decompression layer to the base ranking model.
Assuming the original dimension is $D_{\text{raw}}$ and the compression dimension is $D$, the introduced compression module is denoted as:
\begin{equation}
    h_{\text{compress}} = \sigma\left(\mathcal{W}_1 h + b_1 \right), \label{eq:t_comp}
\end{equation}
where $\mathcal{W}_1 \in \mathcal{R}^{D \times D_{\text{raw}}}$ and $b_1 \in \mathcal{R}^D$ are trainable parameters, and $\sigma$ is the activation function (e.g., ReLU, GELU~\cite{GELU}). $h \in \mathcal{R}^{D_{\text{raw}}}$ denotes the feature representation obtained by Eq.~\ref{eq:fi}, while $h_{\text{compress}} \in \mathcal{R}^{D}$ \textit{is the InsEmb that we aim to save}. The dimension of InsEmb is much smaller than the original dimension, i.e., $D \ll D_{\text{raw}}$.
For effective joint training, the generated InsEmb is projected via an additional lightweight MLP to retain the original dimension:
\begin{equation}
h_{\text{decompress}} = \sigma\left(\mathcal{W}_2 h_{\text{compress}} + b_2\right), \label{eq:t_decomp}
\end{equation}
where $\mathcal{W}_2 \in \mathcal{R}^{D_{\text{raw}} \times D}$ and $b_2 \in \mathcal{R}^{D_{\text{raw}}}$ are trainable parameters. Then, the decompressed feature participates in the final prediction and loss optimization, ensuring that the compressed InsEmb (i.e., $h_{\text{compress}}$) is optimized to capture high-value information. The architecture is illustrated in Fig.~\ref{fig:framework} (a).

\subsubsection{User-Order Source Model}
\label{sec:user_order}
Notably, the above temporal-order source model \textit{compresses training instances individually}, and the obtained InsEmb could only contain the information from a single training instance. Additionally, the compression module may slightly degrade the model performance compared to the base model, which may compromise the quality of the obtained InsEmb. To generate more high-quality InsEmb, we design the \textit{user-order source model} that introduces a \textbf{Source Instance Transformer} (\textbf{SIT}) and enables each instance to perceive historical behaviors of the same user while avoiding future information leakage.

\textbf{User-Order Training Instances.} Differently from the temporal-order one, the user-order source model requires re-organizing training instances to accelerate source model training and InsEmb generation. Typically, training instances of a ranking model are stored and trained according to the action timestamps of all user behaviors. However, this leads to massive redundant storage and computation for the historical behavior sequence items of the same user. During the batch training stage, some recent works propose training models according to user-order instances~\cite{GR,STCA}, which significantly improves efficiency without a performance drop. Specifically, we aggregate training instances within a specific time range by user ID, and then sort the instances of each user according to their temporal order. In practice, the actual batching strategy requires segmenting users’ historical training instances into fixed-length batches according to their counts and optionally padding dummy instances. However, for formula simplicity, we assume that \textit{a batch only contains all the historical training instances of a single user} in the following.

\textbf{Source Instance Transformer.} Assuming we have $T$ historical training instances for a given user, we organize them \textit{in the same batch} and obtain the compressed representations as in Eq.~\ref{eq:t_comp}, which is denoted as $H_{\text{com\_batch}} \in \mathcal{R}^{T \times D}$. In fact, $T$ also denotes the batch size of training instances. Then, we propose the Source Instance Transformer (SIT) to enhance the representation ability of $H_{\text{com\_batch}}$ as follows:
\begin{equation}
    H_{\text{SIT\_batch}} = \mathcal{F}_{\text{transformer}}\left(\text{Reshape}\left(H_{\text{com\_batch}}, [1, T, D]\right), \mathcal{M}\right), \label{eq:u_batch_sit}
\end{equation}
where $\mathcal{F}_{\text{transformer}}$ denotes the common transformer architecture~\cite{Transformer}. The reshape operation transforms the compressed representations of the whole batch (i.e., $H_{\text{com\_batch}}$) into a single sequence with $T$ items. The $T$ items are processed by the transformer module with a causal mask function $\mathcal{M}$, ensuring that each compressed InsEmb can only access the information of past instances. For example, as shown on the left of Fig.~\ref{fig:framework} (b), user instance 1 is the most recent training instance, which could access the information flow from all past training instances. Then, the obtained $H_{\text{SIT\_batch}} \in \mathcal{R}^{1 \times T \times D}$ will first be reshaped to the shape of $[T, D]$ and then decompressed like Eq.~\ref{eq:t_decomp} for final prediction and loss optimization. Notably, we save \textit{the compressed embedding before SIT} (i.e., $H_{\text{com\_batch}}$) as InsEmb instead of $H_{\text{SIT\_batch}}$, which is experimentally verified in Sect.~\ref{sec:ablation}.

Compared with the temporal-order source model, the user-order one has the following advantages:
\begin{itemize}[leftmargin=*]
    \item {\textbf{Enhancing Source Model's Performance}}. SIT enables the information flow between instances within the same user, which could improve the base model's performance significantly as shown in the experimental studies. However, the temporal-order source model may experience a slight performance drop compared with the base model.
    \item {\textbf{Enhancing InsEmb's Sequence Modeling Ability}}. The user-order source model not only generates compact and informative InsEmb, but also implicitly boosts InsEmb's ability of sequence modeling with the help of SIT. In contrast, the temporal-order source model compresses the representations individually for each training instance, losing the latter effect.
\end{itemize}

\begin{figure}[tb]
  \centering
  \includegraphics[width=\linewidth]{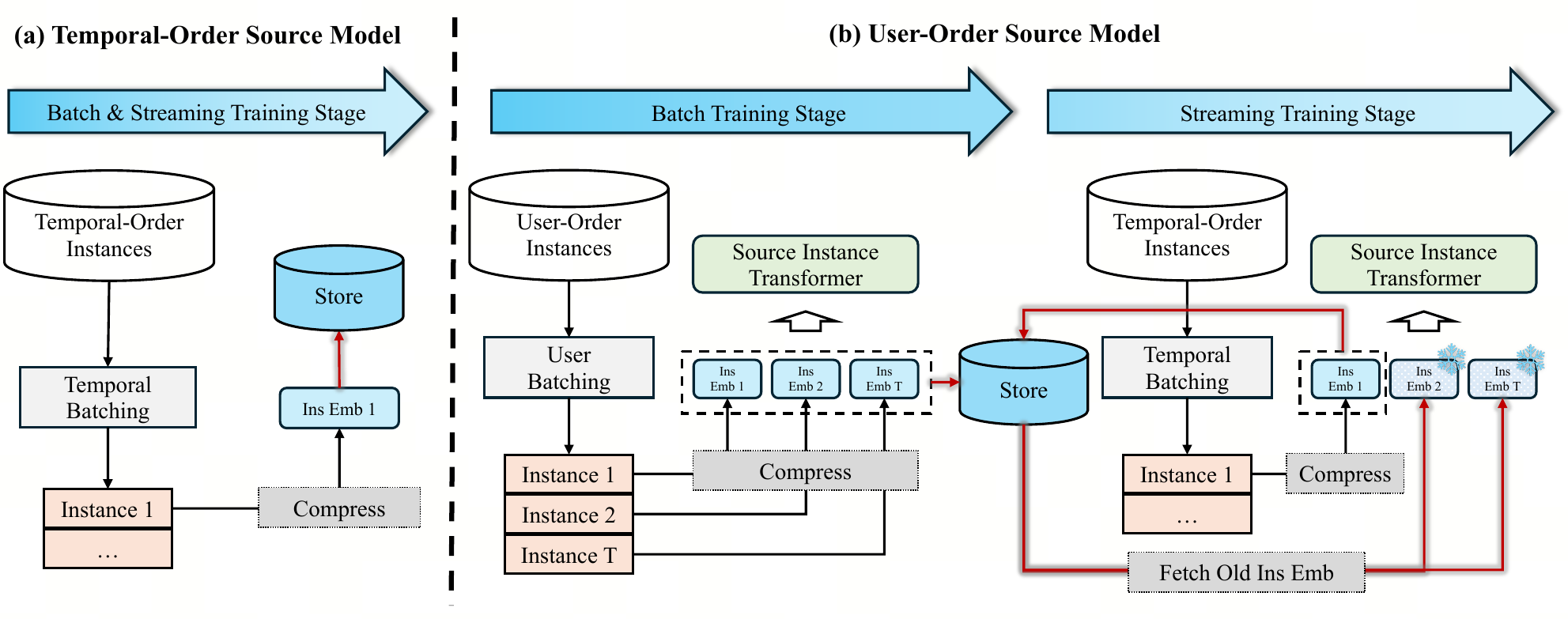}
  \caption{The training paradigms of IAT source models. The temporal-order source model remains consistent between batch and streaming training, while the user-order one requires slight modifications due to data organization change.}
  \label{fig:streaming}
\end{figure}

\subsubsection{Training Paradigms of Source Model}
We adapt a training paradigm that incorporates both batch and stream training stages, where the latter operates on real-time training instances. As shown in Fig.~\ref{fig:streaming}, the temporal-order source model compresses training instances individually, which is irrelevant to the organization of training instances, \textit{keeping consistency between batch and streaming training stages}. However, as declared in Sect.~\ref{sec:user_order}, the user-order source model relies on the user-order training instances during the batch training. During streaming training, the batch of instances commonly comes from various users, and it is difficult to build user-order instances. Hence, we need to \textit{slightly modify the training method of the user-order source model during the streaming stage}.

Specifically, the model architecture keeps unchanged while the batching strategy is different. We denote the compressed representations of a streaming batch as $H_{\text{com\_stream}} \in \mathcal{R}^{B \times D}$, where we use $B$ to denote the batch size. The $B$ instances come from different users. To obtain historical InsEmb required by SIT, we directly retrieve the InsEmb of the previous $T-1$ instances for each user from the store, which is denoted as $H_{\text{com\_stream}}^{\prime} \in \mathcal{R}^{B \times (T-1) \times D}$. Then, the SIT calculation is:
\begin{equation}
    H_{\text{com\_concat}} = \text{Concat}\left(H_{\text{com\_stream}}, \text{sg}(H_{\text{com\_stream}}^{\prime})\right), \label{eq:u_stream_concat}
\end{equation}
\begin{equation}
    H_{\text{SIT\_stream}} = \mathcal{F}_{\text{transformer}}\left(H_{\text{com\_concat}}, \mathcal{M}\right), \label{eq:u_stream_sit}
\end{equation}
where $\text{sg}$ denotes the stop gradient operation, and $H_{\text{com\_concat}} \in \mathcal{R}^{B \times T \times D}$ is the concatenated inputs to the SIT. $T$ is a pre-defined hyperparameter. Finally, we utilize $H_{\text{SIT\_stream}}[:, 0, :] \in \mathcal{R}^{B \times D}$ for the following decompression layer and the following pipelines remain unchanged. In summary, \textit{only the InsEmb of the instances in the streaming batch} (i.e., $H_{\text{com\_stream}}$[:, 0, :]) is newly computed and stored, while the past $T-1$ InsEmb are fetched from the store, which is illustrated in Fig.~\ref{fig:streaming} (b).

\subsection{Stage II: IAT Sequence Modeling}
We detail how to utilize the stored InsEmb to facilitate the downstream tasks in this section. Beyond the informative InsEmb produced by source models, some \textit{core side information} could also be stored and utilized for sequence modeling.

\textbf{Core Side Information}. Although InsEmb contains the compressed information of thousands of features, it does not explicitly cover the information of training instance labels, timestamps, and other task-related information. This module is \textit{optionally} utilized according to practical scenarios. The side information is processed by conventional feature engineering methods, and the embedding representation is denoted as $E_{\text{side}} \in \mathcal{R}^{B \times T \times d_{\text{side}}}$. $B$ is the batch size, and $T$ is the sequence length.

\textbf{InsEmb Adaptation Mechanism}. The stored InsEmb stays in a low-dimensional while compact space, which should be projected to a high-dimensional space for better scaling and feature space matching. Hence, we introduce an adaptation MLP in the downstream model to align InsEmb with the downstream feature space:
\begin{equation}
E_{\text{InsEmb\_adapt}} = \sigma\left(E_{\text{InsEmb}}\mathcal{W}_{\text{adapt}}^T + b_{\text{adapt}}\right), \label{eq:adapt}
\end{equation}
where $\mathcal{W}_{\text{adapt}} \in \mathcal{R}^{d_{\text{adapt}} \times d}$ and $b_{\text{adapt}} \in \mathcal{R}^{d_{\text{adapt}}}$ are the parameters of the adaptation MLP. $E_{\text{InsEmb}} \in \mathcal{R}^{B \times T \times d}$ denotes the fetched historical $T$ InsEmb from the store, whose dimension is $d$.  Commonly, we set $d=D$ and $d_{\text{adapt}} >= d$. 

\textbf{Instance-As-Token (IAT) Sequence}. We construct the \textit{IAT sequence} by concatenating InsEmb and core side information as follows:
\begin{equation}
E_{\text{InsToken}} = \text{Concat}\left(E_{\text{InsEmb\_adapt}}, E_{\text{side}}\right), \label{eq:iat_seq}
\end{equation}
where $E_{\text{InsToken}} \in \mathcal{R}^{B \times T \times (d_{\text{adapt}}+d_{\text{side}})}$ denotes the IAT sequence embedding for sequence modeling. 

\textbf{Query Token Construction}. The IAT sequence is \textit{purely user features}, while modern sequence modeling methods require a query token that contains information of the candidate items for matching prediction. To better align with the compact InsEmb, we concatenate the sequential and non-sequential feature tokens of the downstream model and \textit{compress them as the query}, i.e.,
\begin{equation}
Q = \mathcal{F}_{\text{compress}}\left(\text{Concat}\left(X_{\text{seq}}, X_{\text{non-seq}}\right)\right), \label{eq:query}
\end{equation}
where the obtained query $Q \in \mathcal{R}^{B \times d_{\text{query}}}$ contains enough information. Other query token construction methods are studied in the experiments (i.e., Sect.~\ref{sec:ablation}).

\textbf{IAT Sequence Modeling}. The query token obtained in Eq.~\ref{eq:query} and the IAT sequence obtained in Eq.~\ref{eq:iat_seq} will first be projected to the same dimension space, and then \textit{can be modeled by any mainstream sequence modeling network} according to scenario requirements in the downstream model, including but not limited to LONGER~\cite{LONGER} or Transformer~\cite{Transformer}. We also do some experimental studies to verify the proper position of the modeling results for feature interaction in Sect.~\ref{sec:ablation}. In practice, we find that integrating the modeling results as an input token to the upper complex feature interaction (e.g., RankMixer~\cite{RankMixer}, TokenMixer-Large~\cite{TokenMixerLarge}) obtains the best performance, which is illustrated in Fig.~\ref{fig:framework} (c).

In the streaming training stage, both the source model and downstream model adopt real-time streaming updates. The source model is updated independently, and \textit{the latest InsEmb is synchronized to the downstream model in real time}, ensuring the timeliness of model predictions.

\begin{figure}[tb]
  \centering
  \includegraphics[width=\linewidth]{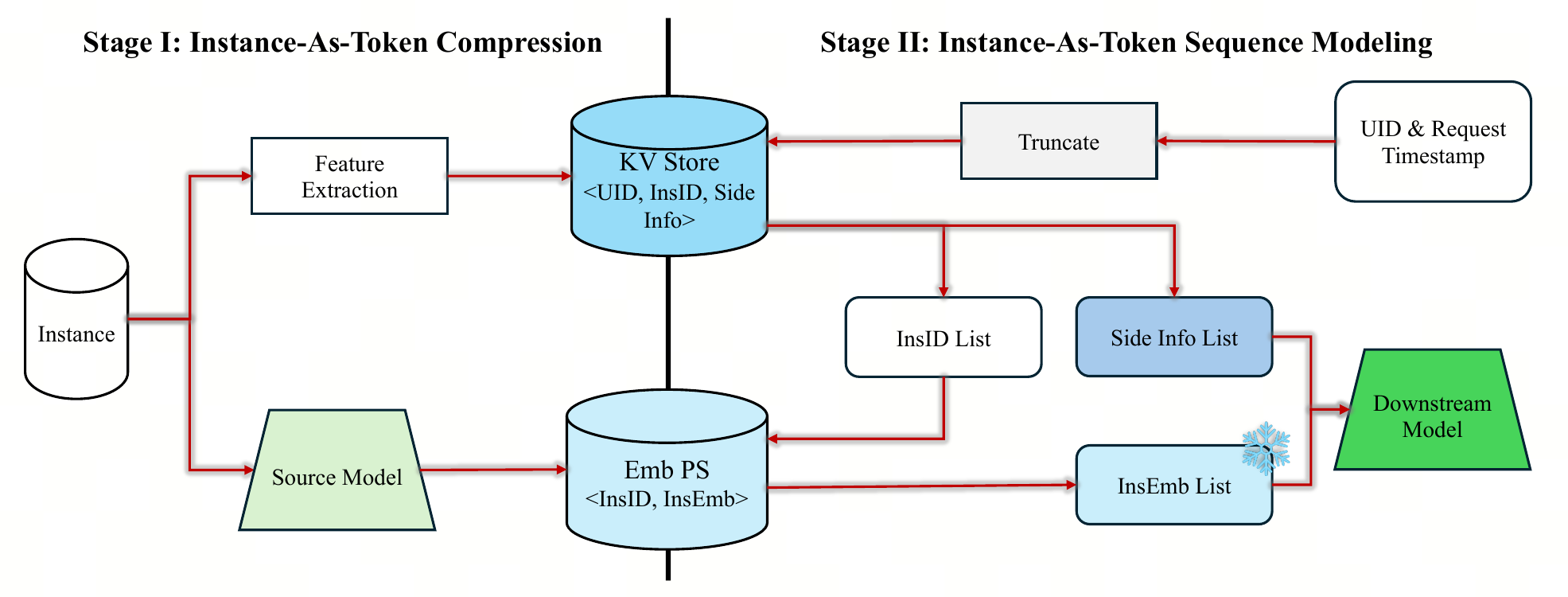}
  \caption{The storage architecture design of IAT. InsID uniquely identifies a training instance.}
  \label{fig:arch}
\end{figure}

\subsection{Storage Architecture Design}
Fig.~\ref{fig:arch} details the storage architecture of IAT, which consists of the following steps: (1) The source model processes training instances and generates the corresponding InsEmb, then writes them to an embedding parameter server (\textbf{Emb PS}) with \textbf{InsID} as the key. (2) Simultaneously, the feature extraction module outputs the tuple of (UID, InsID, Side Information) to a message queue, which is asynchronously saved to a high-performance key-value storage system. \textbf{UID} refers to the hashed user ID. The storage system sorts the InsID sequence by timestamps and truncates it to length $T$ to form the user's historical instance sequence. (3) When a request comes, the downstream model first retrieves the sequence of InsID and Side Info from the high-performance key-value storage system by the key of UID. The sequence is truncated based on the request timestamp and some task-related rules. (4) InsEmb is further retrieved from Emb PS by the key of InsID list, and then the IAT sequence is constructed as introduced in Eq.~\ref{eq:iat_seq} for downstream sequence modeling.

\textbf{Construction of InsID}. InsID is \textit{the unique identifier of a training instance}, which can be constructed by hashing a combination of timestamps and task-related IDs, e.g., the request ID and creative ID in online Ad recommender systems.

\textbf{PS Storage Cost Estimation}.
The storage cost of InsEmb in PS is quantifiable and controllable. Using FP32 (4 bytes per element) for InsEmb storage, the total storage requirement is calculated as:
\begin{equation}
\text{PS Storage Bytes} = N_{\text{daily}} \times T \times d \times 4, \label{eq:storage_bytes}
\end{equation}
where $N_{\text{daily}}$ denotes the average daily number of training instances, $T$ is the number of retention days for historical instances, and $d$ is the InsEmb dimension. This formula accounts for both the daily instance volume and the retention period of historical data, making the storage estimation more consistent with practical industrial scenarios. For example, storing two years of 64-dim InsEmb with approximately hundreds of millions of samples per day consumes several tens of TB of storage space.

\begin{table*}[htb]
\centering
\renewcommand{\arraystretch}{1.2}
\caption{Performance and efficiency comparisons of models without or with IAT. Base models with three distinct sequence modeling architectures for traditional behavior sequences are verified (the leftmost column). Both temporal-order IAT and user-order IAT are compared, and only the results of the downstream models are reported. The downstream models adopt the transformer architecture for the IAT sequence.}
\begin{tabular}{cccccccc}
\toprule
\multicolumn{1}{c}{}         &              & \multicolumn{4}{c}{\textbf{Performance}}          & \multicolumn{2}{c}{\textbf{Efficiency}} \\ 
\cline{3-8}
\multicolumn{1}{c}{}         &              & \textbf{AUC} ($\uparrow$)     & \textbf{$\Delta$AUC (\%)} & \textbf{LogLoss} ($\downarrow$) & \textbf{$\Delta$LogLoss (\%)} & \textbf{Params}     & \textbf{FLOPs/Batch (G, $\times10^9$)}     \\ 
\midrule
\multirow{3}{*}{DIN}         & Base         & 0.83620 & -        & 0.45210 & -            & 49M        &        73G         \\
                             & +Temporal IAT (Trans) & 0.83725 & +0.13\%  & 0.45089 & -0.27\%      & 52M        &      85G           \\ 
                             & \textbf{+User IAT (Trans)} & \textbf{0.83821} & \textbf{+0.24\%}  & \textbf{0.44979} & \textbf{-0.51\%}      & \textbf{52M}        & \textbf{85G}                 \\ 
\midrule
\multirow{3}{*}{LONGER}      & Base         & 0.83705 & -        & 0.45113 & -            & 55M        &         89G        \\
                             & +Temporal IAT (Trans) & 0.83830 & +0.15\%  & 0.44969 & -0.32\%      & 60M        &      97G           \\
                             & \textbf{+User IAT (Trans)} & \textbf{0.83967} & \textbf{+0.31\%}  & \textbf{0.44812} & \textbf{-0.67\%}      & \textbf{60M}        & \textbf{97G}                 \\ 
\midrule
\multirow{3}{*}{Transformer} & Base         & 0.83739 & -        & 0.45075 & -            & 50M        &        126G         \\
                             & +Temporal IAT (Trans) & 0.83868 & +0.15\%  & 0.44928 & -0.33\%      & 50M        &       137G          \\
                             & \textbf{+User IAT (Trans)} & \textbf{0.83984} & \textbf{+0.29\%}  & \textbf{0.44792} & \textbf{-0.63\%}      & \textbf{54M}        & \textbf{137G}                 \\
\bottomrule
\end{tabular}
\label{tab:overall}
\end{table*}

\begin{table}[tb]
\renewcommand{\arraystretch}{1.2}
\caption{Performance comparisons when downstream models take various sequence modeling architectures for the IAT sequence, where we use IAT generated by the user-order source model. The traditional behavior sequence in the base model utilizes the transformer architecture.}
\begin{tabular}{ccccc}
\toprule
\multicolumn{1}{c}{}         & \multicolumn{4}{c}{\textbf{Performance}}\\ 
\cline{2-5}
\multicolumn{1}{c}{}         & \textbf{AUC}   & \textbf{$\Delta$AUC(\%)} & \textbf{LogLoss} & \textbf{$\Delta$LogLoss(\%)}\\ 
\midrule
Base         & 0.8373 & -        & 0.4507 & -            \\
    +IAT (DIN) & 0.8376 & +0.03\%  & 0.4505 & -0.06\%      \\ 
    +IAT (LONGER) & 0.8399 & +0.31\%  & 0.4477 & -0.66\%      \\ 
    +IAT (Trans) & 0.8398 & +0.29\%  & 0.4479 & -0.63\%  \\ 
\bottomrule
\end{tabular}
\label{tab:iat_other_arch}
\end{table}

\section{Experiments}
\subsection{Experimental Setting}
\subsubsection{Datasets}
The experiments are conducted on an industrial-scale CVR prediction dataset collected from a real-world advertising system. The dataset includes 4 consecutive months of training data with a total volume of tens of billion training instances, which is representative of real-world large-scale recommendation scenarios in the advertising industry. The feature space consists user features, item features, contextual features, and traditional behavior sequence features, where the sequence features only retain sparse core side information (e.g., ID, category, action type). All training data are anonymized by removing sensitive user/item information and hashing feature IDs, ensuring no risk of privacy leakage during model training and evaluation.

\subsubsection{Baselines}
The base model architecture conforms to the introduction in Sect.~\ref{sec:base}, while we select three mainstream industrial sequence modeling methods for the traditional behavior sequence (i.e., DIN~\cite{DIN}, LONGER~\cite{LONGER}, Transformer~\cite{Transformer}) to comprehensively validate the effectiveness and flexibility of the IAT sequence. For feature interaction across all models, we uniformly employ RankMixer~\cite{RankMixer} as the feature interaction module to eliminate discrepancies caused by different feature fusion strategies. For each base model, we train the corresponding source model and downstream model with two kinds of IAT sequence (i.e., \textit{Temporal-Order IAT} and \textit{User-Order IAT}), reporting \textit{the AUC gain of them compared to their corresponding base model (only with the traditional hand-crafted behavior sequence)}. In particular, all experiments are trained on a distributed GPU cluster with hundreds of nodes using a batch size of $B=1024$.

\subsubsection{Evaluation Metrics}
To comprehensively evaluate the effectiveness and efficiency of the model, we report both performance metrics and model efficiency metrics. We adopt Area Under Curve (AUC) and Logarithmic Loss (LogLoss) as the performance metrics, which are widely used in industrial CVR prediction tasks. Simultaneously, we report dense parameters and training FLOPs as the primary effiency metrics. We majorly report the performance of downstream models and we also provide the metrics of source models in some cases.

\subsubsection{Hyperparameters}
In the temporal-order source model, the raw dimension of feature representation (i.e., $D_{\text{raw}}$ in Eq.~\ref{eq:t_comp}) is about 6,000 and the compressed dimension is $D=64$. The compressed dimension is the same for the user-order source model, and SIT takes 2 layers of transformer with the model dimension being 64 and the intermediate size of FFN being 128. During the batch training stage of the user-order source model, the input length of SIT is the same as the batch size as shown in Eq.~\ref{eq:u_batch_sit}, while we clip the length of SIT to 256 during the streaming training stage, i.e., the inputs of Eq.~\ref{eq:u_stream_sit} own a shape of $B=1024$, $T=256$ and $D=64$. For the source models, we save the 64-dim compressed representation as InsEmb, which is illustrated as in Fig.~\ref{fig:framework}. The downstream model fetches the InsEmb and adapts it to a dimension of $d_{\text{adapt}}=64$, and we also set the length $T=256$ as shown in Eq.~\ref{eq:adapt}. We set $d_{\text{side}}=64$ and $d_{\text{query}}=512$ for Eq.~\ref{eq:iat_seq} and Eq.~\ref{eq:query}, respectively. Some settings will be slightly changed as we explore different experimental studies in the following. In addition, the length of the traditional behavior sequence is fixed at 512.

\subsection{Overall Performance}
We first comprehensively validated the effectiveness and flexibility of the proposed IAT by integrating it into three mainstream industrial sequential modeling architectures (i.e., DIN~\cite{DIN}, LONGER~\cite{LONGER}, and Transformer~\cite{Transformer}). We apply these architectures to model the traditional behavior sequences in the base model, while only utilizing Transformer for the IAT sequence in the downstream models. In real-world scenarios, these three choices are selected on the basis of available computation resources and constraints. 

\begin{figure}[tb]
  \centering
  \includegraphics[width=\linewidth]{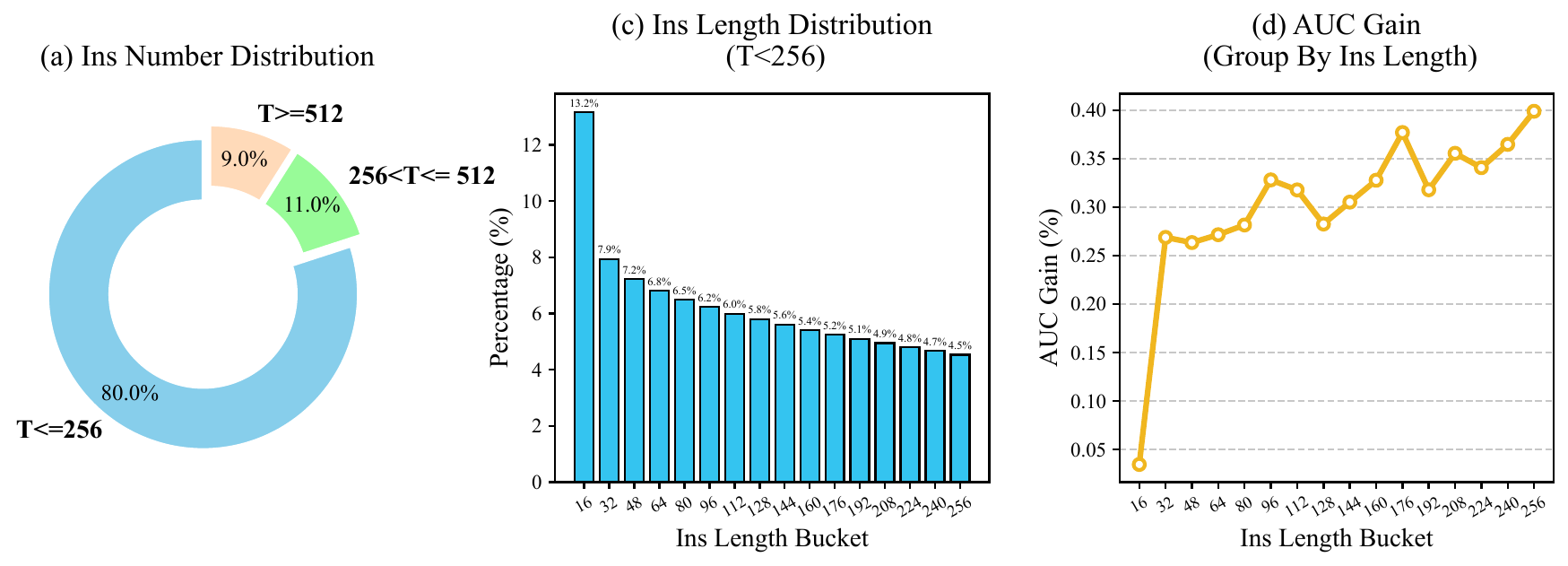}
  \caption{The further analysis of the performance improvements of IAT.}
  \label{fig:iat_stats}
\end{figure}

\subsubsection{The Performance of Source Models}
We separately train a temporal-order and user-order model to generate two types of InsEmb. The temporal-order source model experiences a slight drop in AUC of $0.05\%$, which is the result of the introduced compression module. Meanwhile, the user-order source model obtains a $0.6\%$ AUC gain because the introduced SIT enhances the information flow between historical user instances. However, the user-order source model cannot be directly deployed for serving, as it incurs excessive resource costs and faces practical limitations in online operation.

\subsubsection{The Performance of Downstream Models}
With these two types of InsEmb, we explore their ability to improve the downstream model's performance. The performance and efficiency results of the downstream models are summarized in Tab.~\ref{tab:overall}, where "+Temporal/User IAT" refers to the fact that we add the IAT sequence generated by the temporal-order or user-order source model to the base model.

\begin{table}[tb]
\renewcommand{\arraystretch}{1.2}
\caption{The IAT performance when scaling up the base model. The streaming AUC gains of the source and downstream model over the corresponding base model are reported.}
\begin{tabular}{cccc}
\toprule
\textbf{Scaling Method}         & \textbf{Params}   & \textbf{Source} &  \textbf{Downstream}\\ 
\midrule
Dense Scaling       & 150M & +0.54\% & +0.5\%            \\
Seq. Len. Scaling & 350M  & +0.51\% & +0.30\%      \\ 
Feat. Int. Scaling & 1B & +0.41\% & +0.33\%      \\ 
\bottomrule
\end{tabular}
\label{tab:scale_base}
\end{table}

Across all three sequential architectures, models integrated with the IAT sequence consistently outperform their corresponding base counterparts. Although the temporal-order source model itself experiences a slight performance drop, the benefit to downstream models' AUC is up to $0.15\%$, which verifies the advantage of IAT sequence modeling. In the real-world scene, a $0.1\%$ AUC gain is considered significant. For an intuitive comparison, \textit{introducing a hand-crafted sequence with a length of 256 into the base model can hardly obtain an AUC gain of $0.1\%$} according to our experience. Models with the user-order IAT consistently achieve a relative improvement in AUC (up to $0.31\%$) and reductions in LogLoss (up to $-0.67\%$), accompanied by only a modest increase in parameters and FLOPs. These experiments demonstrate that the IAT sequence is effective in various sequential modeling paradigms.

We observe that \textit{aligning the IAT sequence modeling architecture with the SIT in the user-order source model achieves better results}. Specifically, we fix the traditional behavior sequence in the base model to be modeled by Transformer, and model the IAT sequence with DIN, LONGER, and Transformer, respectively. The LONGER and Transformer lead to significant AUC improvements and LogLoss reductions, as shown in Tab.~\ref{tab:iat_other_arch}, while modeling IAT with DIN only brings weak gains. This indicates that modeling IAT by series of transformer architectures may be more appropriate. In fact, LONGER utilizes the Perceiver~\cite{Perceiver} technique and some other modules (e.g., global token, token merge) to reduce the heavy computation burden of Transformer brought by applying full attention to a long sequence. In the subsequent experiments, unless otherwise specified, both traditional behavior sequences and IAT sequences are modeled using the Transformer architecture.

We further show some analysis on the AUC gain of the user-order IAT. As shown in Fig.~\ref{fig:iat_stats} (a), there are about $80\%$ users who have less than 256 historical training instances. That is why we set the maximum length of the IAT sequence to be 256 for the downstream models, where Fig.~\ref{fig:iat_stats} (b) shows the practical length distribution of the IAT sequence whose length is less than 256. If we group training instances by the length of the IAT, we could find that a longer IAT sequence leads to a more significant AUC gain, which is about $0.4\%$ at most.

\begin{figure}[tb]
  \centering
  \includegraphics[width=0.8\linewidth]{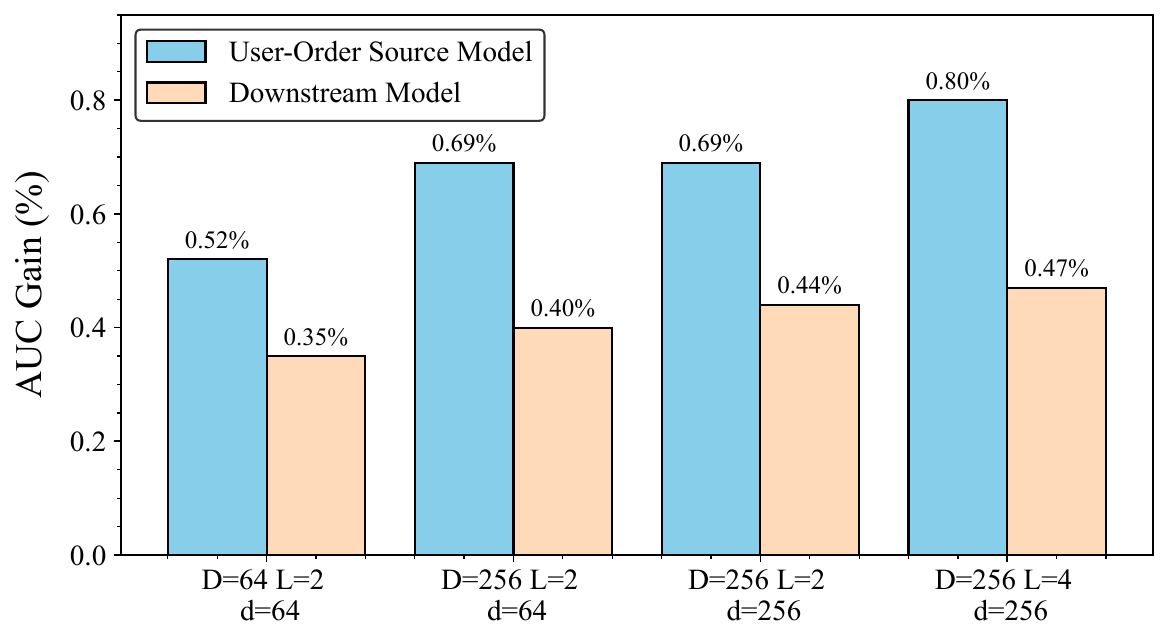}
  \caption{Scaling up results of the SIT in the user-order source model. Both the AUC gains of the source model and downstream model are displayed.}
  \label{fig:scale_sit}
\end{figure}

\begin{figure}[tb]
  \centering
  \includegraphics[width=0.8\linewidth]{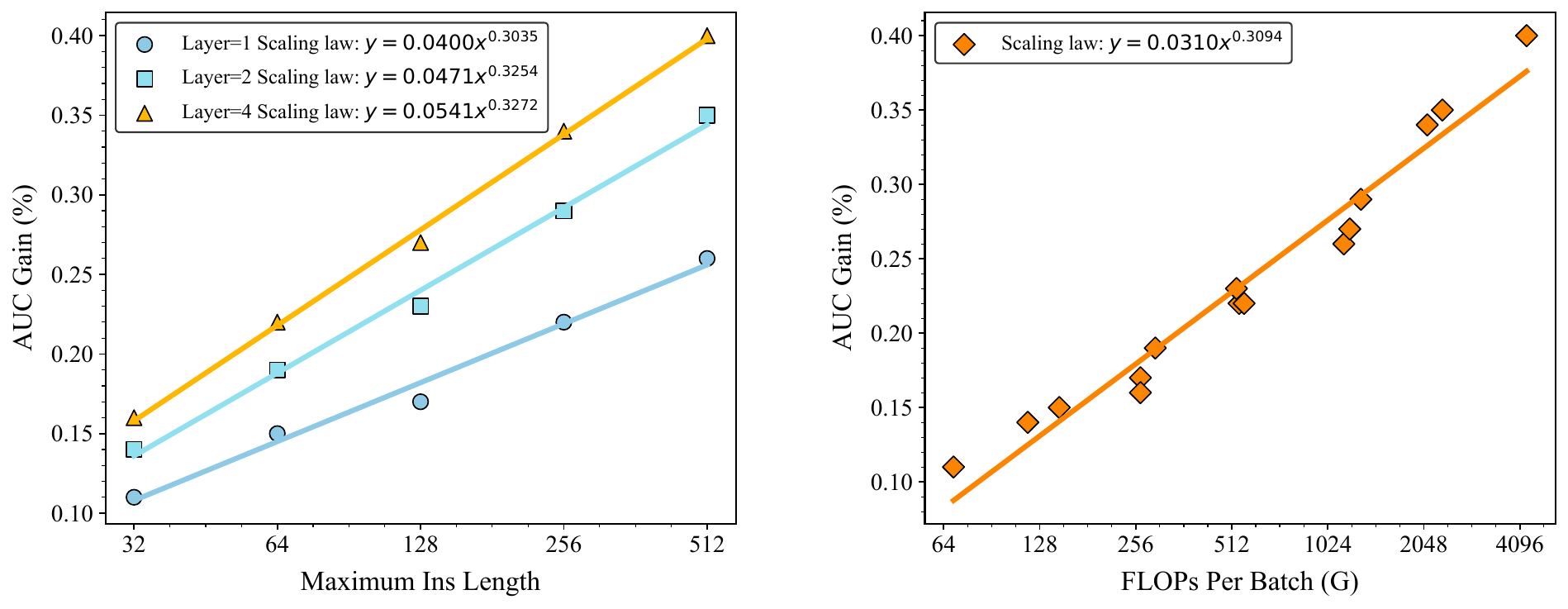}
  \caption{The scaling law of downstream models.}
  \label{fig:scale_law}
\end{figure}

\subsection{Scaling Up Studies}
\subsubsection{Scaling Up the Base Model}
The proposed IAT framework \textit{could be universally applied to various types of base models}. With the development of advanced scaling methods~\cite{Scaling1B,TokenMixerLarge}, we continue to update the online base model in the industrial scene, introducing more dense parameters (Dense Scaling), extending the length of historical behavior sequence (Seq. Len. Scaling), and utilizing more complex feature interaction methods for scaling up (Feat. Int. Scaling). The results are listed in Tab.~\ref{tab:scale_base}. Overall, the downstream model with the user-order IAT sequence obtains a significant performance improvement, especially when the model scales to 1B parameters, for which even a slight gain becomes harder and more valuable.

\subsubsection{Scaling Up the Source Model}
We then explore the scaling up of SIT in the user-order source model. We use $D$, $L$, and $d$ to denote the input dimension of the SIT, the number of transformer layers of SIT, and the finally utilized compressed dimension of InsEmb in downstream models, respectively. The AUC gains over the base model are plotted in Fig.~\ref{fig:scale_sit}. Enlarging the dimension of InsEmb or extending the number of transformer layers brings a significant gain for source models, i.e., an AUC gain from $0.52\%$ to $0.80\%$. The performances of downstream models are also obviously improved, respectively.

\subsubsection{Scaling Up the Downstream Model}
For user-order IAT sequence modeling in the downstream models, we explore the number of transformer layers (i.e., 1, 2, and 4) for different maximum IAT sequence length (i.e., 32, 64, 128, 256, and 512), and plot the scaling law of AUC gain with respect to the maximum IAT length or the FLOPs per batch. The scaling of the transformer architecture of the downstream model conforms to a standard scaling law, where the curves are plotted in Fig.~\ref{fig:scale_law}. Notably, with the number of layers increasing, the downstream model presents a better scaling ability.

\begin{table}[tb]
\renewcommand{\arraystretch}{1.2}
\caption{The ablation studies for IAT. The downstream models' AUC change is reported.}
\begin{tabular}{cccc}
\toprule
 & \textbf{Setting} & \textbf{T-IAT} & \textbf{U-IAT}\\ 
\midrule
\multirow{2}{*}{Source} & Larger $B=2048$ & +0.0\%     & +0.02\%      \\
 & w/ InsEmb Stored after SIT & -  & -0.09\%      \\
\midrule
\multirow{4}{*}{Downstream} & w/o label sideinfo & -0.06\%     & -0.10\%      \\
 & w/o other sideinfo & -0.05\%  & -0.01\%      \\
 & w/ ID-based query & -0.06\%  & -0.09\%      \\
 & w/ IAT after Feat. Int. & -0.04\%  & -0.06\%      \\
\bottomrule
\end{tabular}
\label{tab:ablation}
\end{table}

\subsection{Ablation Studies}
\label{sec:ablation}
Then, we perform ablation studies in Tab.~\ref{tab:ablation} for some settings of the IAT, including those for source models and downstream models, respectively. For source models, we first change the batch size $B$ from $1024$ to $2048$, which has no effect for the temporal-order IAT (T-IAT), but slightly enhances the performance of the user-order IAT (U-IAT) because \textit{a batch could cover more historical instances of each user during the user-order batch training stage}. Saving the InsEmb after SIT module instead of those before SIT reduces the downstream performance, because the InsEmb after SIT loses some fundamental information and becomes more similar due to the attention mechanism. The IAT sequence contains both InsEmb and some core side information (sideinfo) such as the multitask labels that represent the behavior actions and some other task-related side information. We use 5 other types of side information in practice. Dropping the label side information leads to an obvious AUC degradation, because the InsEmb contains few information about the labels of training instances. Dropping other sideinfo has a bad effect on the temporal-order IAT, while it does not impact the performance of the user-order one. We also try using a simpler query constructed by some ID features of a candidate instead of compressing the sequential and non-sequential tokens (Eq.~\ref{eq:query}), but the resulted AUC drops $0.06\%$ and $0.09\%$, respectively. Finally, we observe that the position of IAT sequence modeling in the downstream models also matters, and place it before the feature interaction module (e.g., RankMixer) acquires a better result. More details of ablation studies are listed in the Appendix.

\begin{table}[tb]
\renewcommand{\arraystretch}{1.2}
\caption{Online A/B results on in-domain advertising scenes.}
\begin{tabular}{ccc}
\toprule
 & \textbf{ADSS} & \textbf{ADVV} \\ 
\midrule
Temporal-Order IAT & +0.685\%     & +0.653\%      \\
User-Order IAT & +1.557\%& +1.340\%      \\
\bottomrule
\end{tabular}
\label{tab:in_domain_ab}
\end{table}

\begin{table}[tb]
\renewcommand{\arraystretch}{1.2}
\caption{Online A/B results on other cross-domain scenes.}
\begin{tabular}{cccc}
\toprule
\textbf{Scene} & \textbf{Model} & \textbf{Metric} & \textbf{Uplift} \\ 
\midrule
Mall Advertising & CVR     & ADVV  &  +1.482\%   \\
Feed Advertising & CVR & ADSS   & +0.5\%  \\
Feed Advertising & CTR & ADSS   & +3.015\%  \\
NOC Advertising & CVR & ADSS   & +1.19\%  \\
Live Streaming E-Com. & CT-CVR & GMV   & +0.151\%  \\
\bottomrule
\end{tabular}
\label{tab:cross_domain_ab}
\end{table}

\subsection{Online A/B Results}
We generate the IAT sequences by training both temporal-order and user-order source models on \textit{a real-world advertising scene}. Then we apply the produced IAT sequence to \textit{the advertising scene itself} and \textit{several other industrial recommendation scenes}.
\subsubsection{In-Domain A/B Results}
Applying the IAT sequence to the advertising scene itself shows the in-domain transferability of the proposed IAT. Core revenue metrics including ADSS (Advertiser Score) and ADVV (Advertiser Value) are reported in Tab.~\ref{tab:in_domain_ab}. The base is \textit{a strong advertising ranking model that has served online for a long time}. Clearly, IAT delivers statistically significant uplifts on those metrics.

\subsubsection{Cross-Domain A/B Results}
The IAT also demonstrates strong generalization ability \textit{when applying the produced IAT sequence to other downstream scenarios}, including the CVR prediction of mall advertising, the CTR and CVR prediction of feed advertising, the CVR prediction of non-closed-loop advertising, and the CT-CVR prediction of the live-streaming e-commerce. The core metrics for advertising include ADSS and ADVV, while the GMV is reported for the e-commerce scenes. As listed in Tab.~\ref{tab:cross_domain_ab}, IAT obtains significant uplifts across all of these scenes. 

\section{Conclusion}
Instead of conventional hand-crafted sequence engineering, we propose a novel two-stage sequence modeling framework named \textbf{Instance-As-Token} (\textbf{IAT}). The first stage of IAT constructs either a temporal-order or user-order source model to compress the abundant features of each user’s historical instances into compact and informative embeddings (i.e., InsEmb). Then, the downstream stage utilizes these InsEmb for state-of-the-art sequence modeling. Extensive offline and online experiments demonstrate the superiority of the proposed IAT in improving model quality and business-critical metrics. The proposed IAT has been fully deployed in several industrial recommender systems. Future work will further improve the efficiency of IAT, including: (1) exploring a one-stage framework to reduce training complexity; (2) adopting more advanced compression techniques to further decrease storage and transmission overhead.

\bibliographystyle{ACM-Reference-Format}
\bibliography{iat}

@String{Computing = "Computing" }

@inproceedings{TWINV1,
  title={TWIN: TWo-stage interest network for lifelong user behavior modeling in CTR prediction at kuaishou},
  author={Chang, Jianxin and Zhang, Chenbin and Fu, Zhiyi and Zang, Xiaoxue and Guan, Lin and Lu, Jing and Hui, Yiqun and Leng, Dewei and Niu, Yanan and Song, Yang and others},
  booktitle={Proceedings of the 29th ACM SIGKDD Conference on Knowledge Discovery and Data Mining},
  pages={3785--3794},
  year={2023}
}

@inproceedings{TWINV2,
  title={Twin v2: Scaling ultra-long user behavior sequence modeling for enhanced ctr prediction at kuaishou},
  author={Si, Zihua and Guan, Lin and Sun, ZhongXiang and Zang, Xiaoxue and Lu, Jing and Hui, Yiqun and Cao, Xingchao and Yang, Zeyu and Zheng, Yichen and Leng, Dewei and others},
  booktitle={Proceedings of the 33rd ACM International Conference on Information and Knowledge Management},
  pages={4890--4897},
  year={2024}
}

@inproceedings{TransActV2,
  title={TransAct V2: Lifelong User Action Sequence Modeling on Pinterest Recommendation},
  author={Xia, Xue and Joshi, Saurabh and Rajesh, Kousik and Li, Kangnan and Lu, Yangyi and Pancha, Nikil and Badani, Dhruvil and Xu, Jiajing and Eksombatchai, Pong},
  booktitle={Proceedings of the 34th ACM International Conference on Information and Knowledge Management},
  pages={6881--6882},
  year={2025}
}

@article{ENCODE,
  title={ENCODE: Breaking the trade-off between performance and efficiency in long-term user behavior modeling},
  author={Zhou, Wen-Ji and Zheng, Yuhang and Feng, Yinfu and Ye, Yunan and Xiao, Rong and Chen, Long and Yang, Xiaosong and Xiao, Jun},
  journal={IEEE Transactions on Knowledge and Data Engineering},
  volume={37},
  number={1},
  pages={265--277},
  year={2024},
}

@inproceedings{LREA,
  title={Lrea: Low-rank efficient attention on modeling long-term user behaviors for ctr prediction},
  author={Song, Xin and Li, Xiaochen and Hu, Jinxin and Wen, Hong and Chen, Zulong and Zhang, Yu and Zeng, Xiaoyi and Zhang, Jing},
  booktitle={Proceedings of the 48th International ACM SIGIR Conference on Research and Development in Information Retrieval},
  pages={2843--2847},
  year={2025}
}

@article{VQL,
  title={VQL: An End-to-End Context-Aware Vector Quantization Attention for Ultra-Long User Behavior Modeling},
  author={Li, Kaiyuan and Tang, Yongxiang and Cheng, Yanhua and Bai, Yong and Zeng, Yanxiang and Wang, Chao and Liu, Xialong and Jiang, Peng},
  journal={arXiv preprint arXiv:2508.17125},
  year={2025}
}

@inproceedings{DMQN,
  title={Deep Multiple Quantization Network on Long Behavior Sequence for Click-Through Rate Prediction},
  author={Wei, Zhuoxing and Liu, Qi and Xie, Qingchen},
  booktitle={Proceedings of the 48th International ACM SIGIR Conference on Research and Development in Information Retrieval},
  pages={3090--3094},
  year={2025}
}

@inproceedings{DV365,
  title={DV365: Extremely Long User History Modeling at Instagram},
  author={Lyu, Wenhan and Tyagi, Devashish and Yang, Yihang and Li, Ziwei and Somani, Ajay and Shanmugasundaram, Karthikeyan and Andrejevic, Nikola and Adeputra, Ferdi and Zeng, Curtis and Singh, Arun K and others},
  booktitle={Proceedings of the 31st ACM SIGKDD Conference on Knowledge Discovery and Data Mining V. 2},
  pages={4717--4727},
  year={2025}
}

@inproceedings{LONGER,
  title={Longer: Scaling up long sequence modeling in industrial recommenders},
  author={Chai, Zheng and Ren, Qin and Xiao, Xijun and Yang, Huizhi and Han, Bo and Zhang, Sijun and Chen, Di and Lu, Hui and Zhao, Wenlin and Yu, Lele and others},
  booktitle={Proceedings of the Nineteenth ACM Conference on Recommender Systems},
  pages={247--256},
  year={2025}
}

@article{STCA,
  title={Make It Long, Keep It Fast: End-to-End 10k-Sequence Modeling at Billion Scale on Douyin},
  author={Guan, Lin and Yang, Jia-Qi and Zhao, Zhishan and Zhang, Beichuan and Sun, Bo and Luo, Xuanyuan and Ni, Jinan and Li, Xiaowen and Qi, Yuhang and Fan, Zhifang and others},
  journal={arXiv preprint arXiv:2511.06077},
  year={2025}
}

@article{HiSAC,
  title={HiSAC: Hierarchical Sparse Activation Compression for Ultra-long Sequence Modeling in Recommenders},
  author={Yuan, Kun and Bi, Junyu and Cheng, Daixuan and Wu, Changfa and Xiao, Shuwen and Cao, Binbin and Wu, Jian and Jiang, Yuning},
  journal={arXiv preprint arXiv:2602.21009},
  year={2026}
}

@article{HiFormer,
  title={Hiformer: Heterogeneous feature interactions learning with transformers for recommender systems},
  author={Gui, Huan and Wang, Ruoxi and Yin, Ke and Jin, Long and Kula, Maciej and Xu, Taibai and Hong, Lichan and Chi, Ed H},
  journal={arXiv preprint arXiv:2311.05884},
  year={2023}
}

@inproceedings{InterFormer,
  title={InterFormer: Effective Heterogeneous Interaction Learning for Click-Through Rate Prediction},
  author={Zeng, Zhichen and Liu, Xiaolong and Hang, Mengyue and Liu, Xiaoyi and Zhou, Qinghai and Yang, Chaofei and Liu, Yiqun and Ruan, Yichen and Chen, Laming and Chen, Yuxin and others},
  booktitle={Proceedings of the 34th ACM International Conference on Information and Knowledge Management},
  pages={6225--6233},
  year={2025}
}

@article{OneTrans,
  title={OneTrans: Unified Feature Interaction and Sequence Modeling with One Transformer in Industrial Recommender},
  author={Zhang, Zhaoqi and Pei, Haolei and Guo, Jun and Wang, Tianyu and Feng, Yufei and Sun, Hui and Liu, Shaowei and Sun, Aixin},
  journal={arXiv preprint arXiv:2510.26104},
  year={2025}
}

@article{HyFormer,
  title={HyFormer: Revisiting the Roles of Sequence Modeling and Feature Interaction in CTR Prediction},
  author={Huang, Yunwen and Hong, Shiyong and Xiao, Xijun and Jin, Jinqiu and Luo, Xuanyuan and Wang, Zhe and Chai, Zheng and Wu, Shikang and Zheng, Yuchao and Lin, Jingjian},
  journal={arXiv preprint arXiv:2601.12681},
  year={2026}
}

@article{MUSE,
  title={MUSE: A Simple Yet Effective Multimodal Search-Based Framework for Lifelong User Interest Modeling},
  author={Wu, Bin and Yang, Feifan and Chan, Zhangming and Gu, Yu-Ran and Feng, Jiawei and Yi, Chao and Sheng, Xiang-Rong and Zhu, Han and Xu, Jian and Ye, Mang and others},
  journal={arXiv preprint arXiv:2512.07216},
  year={2025}
}

@article{LEMUR,
  title={LEMUR: Large scale End-to-end MUltimodal Recommendation},
  author={Han, Xintian and Chen, Honggang and Lin, Quan and Gao, Jingyue and Ren, Xiangyuan and Zhu, Lifei and Ye, Zhisheng and Wu, Shikang and Xie, XiongHang and Gan, Xiaochu and others},
  journal={arXiv preprint arXiv:2511.10962},
  year={2025}
}

@article{Scaling1B,
  title={Scaling recommender transformers to one billion parameters},
  author={Khrylchenko, Kirill and Matveev, Artem and Makeev, Sergei and Baikalov, Vladimir},
  journal={arXiv preprint arXiv:2507.15994},
  year={2025}
}

@inproceedings{CLIMBER,
  title={Climber: Toward efficient scaling laws for large recommendation models},
  author={Xu, Songpei and Wang, Shijia and Guo, Da and Guo, Xianwen and Xiao, Qiang and Huang, Bin and Wu, Guanlin and Luo, Chuanjiang},
  booktitle={Proceedings of the 34th ACM International Conference on Information and Knowledge Management},
  pages={6193--6200},
  year={2025}
}

@inproceedings{SUAN,
  title={Exploring Scaling Laws of CTR Model for Online Performance Improvement},
  author={Lai, Weijiang and Jin, Beihong and Zhang, Jiongyan and Zheng, Yiyuan and Dong, Jian and Cheng, Jia and Lei, Jun and Wang, Xingxing},
  booktitle={Proceedings of the Nineteenth ACM Conference on Recommender Systems},
  pages={114--123},
  year={2025}
}

@article{Wukong,
  title={Wukong: Towards a scaling law for large-scale recommendation},
  author={Zhang, Buyun and Luo, Liang and Chen, Yuxin and Nie, Jade and Liu, Xi and Guo, Daifeng and Zhao, Yanli and Li, Shen and Hao, Yuchen and Yao, Yantao and others},
  journal={arXiv preprint arXiv:2403.02545},
  year={2024}
}

@inproceedings{RankMixer,
  title={Rankmixer: Scaling up ranking models in industrial recommenders},
  author={Zhu, Jie and Fan, Zhifang and Zhu, Xiaoxie and Jiang, Yuchen and Wang, Hangyu and Han, Xintian and Ding, Haoran and Wang, Xinmin and Zhao, Wenlin and Gong, Zhen and others},
  booktitle={Proceedings of the 34th ACM International Conference on Information and Knowledge Management},
  pages={6309--6316},
  year={2025}
}

@article{TokenMixerLarge,
  title={TokenMixer-Large: Scaling Up Large Ranking Models in Industrial Recommenders},
  author={Jiang, Yuchen and Zhu, Jie and Han, Xintian and Lu, Hui and Bai, Kunmin and Yang, Mingyu and Wu, Shikang and Zhang, Ruihao and Zhao, Wenlin and Bai, Shipeng and others},
  journal={arXiv preprint arXiv:2602.06563},
  year={2026}
}

@article{HLLM,
  title={Hllm: Enhancing sequential recommendations via hierarchical large language models for item and user modeling},
  author={Chen, Junyi and Chi, Lu and Peng, Bingyue and Yuan, Zehuan},
  journal={arXiv preprint arXiv:2409.12740},
  year={2024}
}

@inproceedings{MARM,
  title={MARM: Unlocking the Recommendation Cache Scaling-Law through Memory Augmentation and Scalable Complexity},
  author={Lv, Xiao and Cao, Jiangxia and Guan, Shijie and Zhou, Xiaoyou and Qi, Zhiguang and Zang, Yaqiang and Wang, Ben and Zhou, Guorui},
  booktitle={Proceedings of the 34th ACM International Conference on Information and Knowledge Management},
  pages={2022--2031},
  year={2025}
}

@article{LargeFM,
  title={External Large Foundation Model: How to Efficiently Serve Trillions of Parameters for Online Ads Recommendation},
  author={Recommendation, Ads},
  journal={arXiv preprint arXiv:2502.17494},
  year={2025}
}

@article{LLaTTE,
  title={LLaTTE: Scaling Laws for Multi-Stage Sequence Modeling in Large-Scale Ads Recommendation},
  author={Xiong, Lee and Chen, Zhirong and Mayuranath, Rahul and Qiu, Shangran and Ozdemir, Arda and Li, Lu and Hu, Yang and Li, Dave and Ren, Jingtao and Cheng, Howard and others},
  journal={arXiv preprint arXiv:2601.20083},
  year={2026}
}

@article{GR,
  title={Actions speak louder than words: Trillion-parameter sequential transducers for generative recommendations},
  author={Zhai, Jiaqi and Liao, Lucy and Liu, Xing and Wang, Yueming and Li, Rui and Cao, Xuan and Gao, Leon and Gong, Zhaojie and Gu, Fangda and He, Michael and others},
  journal={arXiv preprint arXiv:2402.17152},
  year={2024}
}

@article{TIGER,
  title={Recommender systems with generative retrieval},
  author={Rajput, Shashank and Mehta, Nikhil and Singh, Anima and Hulikal Keshavan, Raghunandan and Vu, Trung and Heldt, Lukasz and Hong, Lichan and Tay, Yi and Tran, Vinh and Samost, Jonah and others},
  journal={Advances in Neural Information Processing Systems},
  volume={36},
  pages={10299--10315},
  year={2023}
}

@article{OneRec,
  title={Onerec: Unifying retrieve and rank with generative recommender and iterative preference alignment},
  author={Deng, Jiaxin and Wang, Shiyao and Cai, Kuo and Ren, Lejian and Hu, Qigen and Ding, Weifeng and Luo, Qiang and Zhou, Guorui},
  journal={arXiv preprint arXiv:2502.18965},
  year={2025}
}

@article{Transformer,
  title={Attention is all you need},
  author={Vaswani, Ashish and Shazeer, Noam and Parmar, Niki and Uszkoreit, Jakob and Jones, Llion and Gomez, Aidan N and Kaiser, {\L}ukasz and Polosukhin, Illia},
  journal={Advances in neural information processing systems},
  volume={30},
  year={2017}
}

@inproceedings{DIN,
  title={Deep interest evolution network for click-through rate prediction},
  author={Zhou, Guorui and Mou, Na and Fan, Ying and Pi, Qi and Bian, Weijie and Zhou, Chang and Zhu, Xiaoqiang and Gai, Kun},
  booktitle={Proceedings of the AAAI conference on artificial intelligence},
  volume={33},
  number={01},
  pages={5941--5948},
  year={2019}
}

@inproceedings{UniSRec,
  title={Towards universal sequence representation learning for recommender systems},
  author={Hou, Yupeng and Mu, Shanlei and Zhao, Wayne Xin and Li, Yaliang and Ding, Bolin and Wen, Ji-Rong},
  booktitle={Proceedings of the 28th ACM SIGKDD conference on knowledge discovery and data mining},
  pages={585--593},
  year={2022}
}

@inproceedings{FeatureTutorial,
  title={Tutorial: feature engineering for recommender systems},
  author={Schifferer, Benedikt and Deotte, Chris and Oldridge, Even},
  booktitle={Proceedings of the 14th ACM Conference on Recommender Systems},
  pages={754--755},
  year={2020}
}

@inproceedings{PretrainUE,
  title={Augmenting sequential recommendation with pseudo-prior items via reversely pre-training transformer},
  author={Liu, Zhiwei and Fan, Ziwei and Wang, Yu and Yu, Philip S},
  booktitle={Proceedings of the 44th international ACM SIGIR conference on Research and development in information retrieval},
  pages={1608--1612},
  year={2021}
}

@inproceedings{NSA,
  title={Native sparse attention: Hardware-aligned and natively trainable sparse attention},
  author={Yuan, Jingyang and Gao, Huazuo and Dai, Damai and Luo, Junyu and Zhao, Liang and Zhang, Zhengyan and Xie, Zhenda and Wei, Yuxing and Wang, Lean and Xiao, Zhiping and others},
  booktitle={Proceedings of the 63rd Annual Meeting of the Association for Computational Linguistics (Volume 1: Long Papers)},
  pages={23078--23097},
  year={2025}
}

@article{ScalingLaw,
  title={Scaling laws for neural language models},
  author={Kaplan, Jared and McCandlish, Sam and Henighan, Tom and Brown, Tom B and Chess, Benjamin and Child, Rewon and Gray, Scott and Radford, Alec and Wu, Jeffrey and Amodei, Dario},
  journal={arXiv preprint arXiv:2001.08361},
  year={2020}
}

@inproceedings{MTGR,
  title={Mtgr: Industrial-scale generative recommendation framework in meituan},
  author={Han, Ruidong and Yin, Bin and Chen, Shangyu and Jiang, He and Jiang, Fei and Li, Xiang and Ma, Chi and Huang, Mincong and Li, Xiaoguang and Jing, Chunzhen and others},
  booktitle={Proceedings of the 34th ACM International Conference on Information and Knowledge Management},
  pages={5731--5738},
  year={2025}
}

@article{OneMall,
  title={OneMall: One Model, More Scenarios--End-to-End Generative Recommender Family at Kuaishou E-Commerce},
  author={Zhang, Kun and Zhang, Jingming and Cheng, Wei and Cheng, Yansong and Zhang, Jiaqi and Lu, Hao and Zhang, Xu and Gan, Haixiang and Cao, Jiangxia and Wang, Tenglong and others},
  journal={arXiv preprint arXiv:2601.21770},
  year={2026}
}

@inproceedings{Rabbitail,
  title={Rabbitail: A Tail Latency-Aware Scheduler for Deep Learning Recommendation Systems with Hierarchical Embedding Storage},
  author={Wan, Hu and Huang, Yun and Bai, Shuhan and Sun, Xuan and Kuo, Tei-Wei and Xue, Chun Jason},
  booktitle={Proceedings of the 40th ACM/SIGAPP Symposium on Applied Computing},
  pages={279--287},
  year={2025}
}

@article{SCRec,
  title={SCRec: A Scalable Computational Storage System with Statistical Sharding and Tensor-train Decomposition for Recommendation Models},
  author={Yang, Jinho and Kim, Ji-Hoon and Kim, Joo-Young},
  journal={IEEE Transactions on Computers},
  year={2025},
  publisher={IEEE}
}

@inproceedings{ASIF,
  title={Aligned side information fusion method for sequential recommendation},
  author={Wang, Shuhan and Shen, Bin and Min, Xu and He, Yong and Zhang, Xiaolu and Zhang, Liang and Zhou, Jun and Mo, Linjian},
  booktitle={Companion Proceedings of the ACM Web Conference 2024},
  pages={112--120},
  year={2024}
}

@inproceedings{SASRec,
  title={Self-attentive sequential recommendation},
  author={Kang, Wang-Cheng and McAuley, Julian},
  booktitle={2018 IEEE international conference on data mining (ICDM)},
  pages={197--206},
  year={2018},
  organization={IEEE}
}

@inproceedings{FDSA,
  title={Feature-level deeper self-attention network for sequential recommendation},
  author={Zhang, Tingting and Zhao, Pengpeng and Liu, Yanchi and Sheng, Victor S and Xu, Jiajie and Wang, Deqing and Liu, Guanfeng and Zhou, Xiaofang and others},
  booktitle={IJCAI},
  pages={4320--4326},
  year={2019}
}

@inproceedings{DIF,
  title={Decoupled side information fusion for sequential recommendation},
  author={Xie, Yueqi and Zhou, Peilin and Kim, Sunghun},
  booktitle={Proceedings of the 45th international ACM SIGIR conference on research and development in information retrieval},
  pages={1611--1621},
  year={2022}
}

@inproceedings{Trans2D,
  title={Sequential modeling with multiple attributes for watchlist recommendation in e-commerce},
  author={Singer, Uriel and Roitman, Haggai and Eshel, Yotam and Nus, Alexander and Guy, Ido and Levi, Or and Hasson, Idan and Kiperwasser, Eliyahu},
  booktitle={Proceedings of the fifteenth ACM international conference on web search and data mining},
  pages={937--946},
  year={2022}
}

@article{FORGE,
  title={Forge: Forming semantic identifiers for generative retrieval in industrial datasets},
  author={Fu, Kairui and Zhang, Tao and Xiao, Shuwen and Wang, Ziyang and Zhang, Xinming and Zhang, Chenchi and Yan, Yuliang and Zheng, Junjun and Li, Yu and Chen, Zhihong and others},
  journal={arXiv preprint arXiv:2509.20904},
  year={2025}
}

@article{UBMSurvey,
  title={A survey on user behavior modeling in recommender systems},
  author={He, Zhicheng and Liu, Weiwen and Guo, Wei and Qin, Jiarui and Zhang, Yingxue and Hu, Yaochen and Tang, Ruiming},
  journal={arXiv preprint arXiv:2302.11087},
  year={2023}
}

@inproceedings{ERASE,
  title={Erase: Benchmarking feature selection methods for deep recommender systems},
  author={Jia, Pengyue and Wang, Yejing and Du, Zhaocheng and Zhao, Xiangyu and Wang, Yichao and Chen, Bo and Wang, Wanyu and Guo, Huifeng and Tang, Ruiming},
  booktitle={Proceedings of the 30th ACM SIGKDD conference on knowledge discovery and data mining},
  pages={5194--5205},
  year={2024}
}

@article{GELU,
  title={Gaussian error linear units (gelus)},
  author={Hendrycks, Dan and Gimpel, Kevin},
  journal={arXiv preprint arXiv:1606.08415},
  year={2016}
}

@inproceedings{Perceiver,
  title={Perceiver: General perception with iterative attention},
  author={Jaegle, Andrew and Gimeno, Felix and Brock, Andy and Vinyals, Oriol and Zisserman, Andrew and Carreira, Joao},
  booktitle={International conference on machine learning},
  pages={4651--4664},
  year={2021},
  organization={PMLR}
}

\clearpage
\appendix

\section{Appendix}

\subsection{Details of the Ablation Studies}
We detail the ablation studies introduced in the body text as follows. In our scene, we use some side information in addition to the InsEmb, formulating the InsToken for IAT sequence modeling. The side information contains in general two parts:
\begin{itemize}[leftmargin=*]
    \item {\textbf{Label Side Information}}: the multi-task labels of a user's historical training instance, e.g., whether the user bought the clicked item or not. These labels are highly relevant to the user behaviors, while the InsEmb only compresses the bottom feature representations. Hence, we regard the label side information as important features and could be seamlessly combined with the InsEmb for better sequence modeling.
    \item {\textbf{Other Task-Related Side Information}}: in our advertising scene, several types of timestamps that could describe the attribution logic of a user's behavior are also important for sequence modeling, and we also consider those side information.
\end{itemize}
 The side information is optional according to the real-world scenarios, and the side information could be represented by ID features and modeled by embedding tables in the model. Only the hashed values of such side information are stored and utilized in our model.

\begin{figure}[b]
  \centering
  \includegraphics[width=\linewidth]{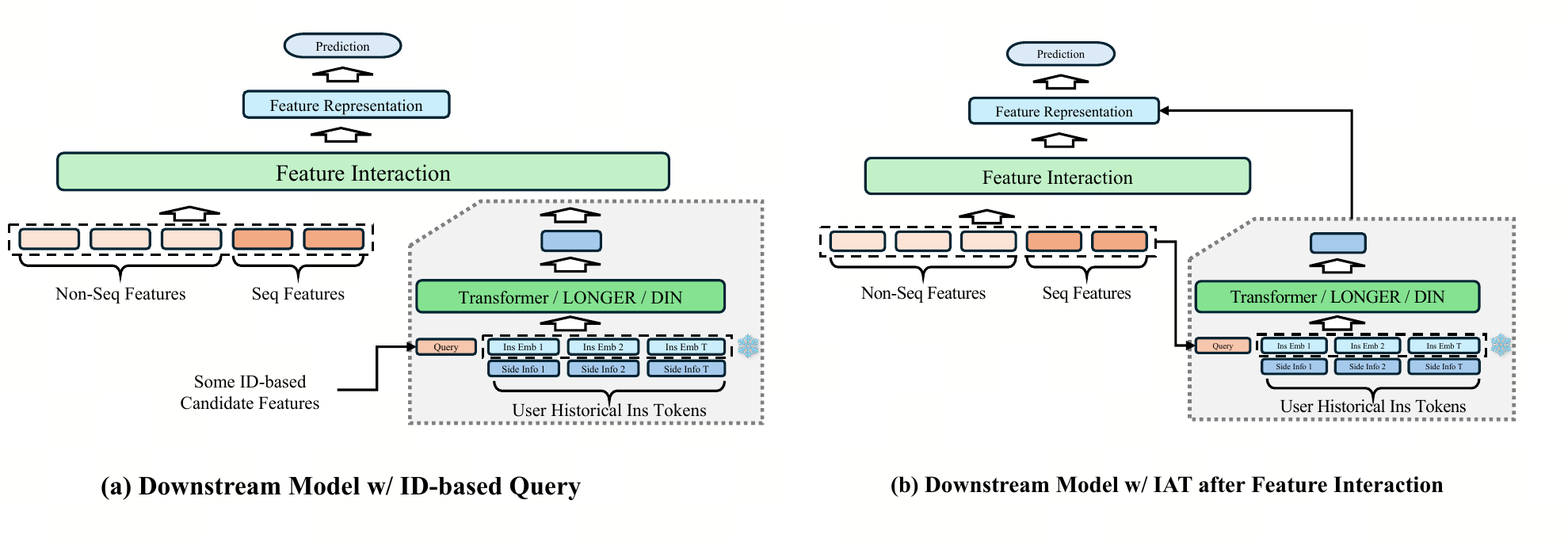}
  \caption{Different IAT modeling architecture choices studied in the ablation studies.}
  \label{fig:ablation}
\end{figure}

 Then, we show two other varieties of IAT sequence modeling, which study the importance of query construction and IAT position, correspondingly. These two choices are shown in Fig.~\ref{fig:ablation}. However, these two architectures suffer obvious performance degradation. First, because the InsToken in the IAT sequence contains informative information, and hence it is better to use a query token that also covers informative features of the candidate. Second, placing the output token before the complex feature interaction module could result in a comprehensive information flow among various bottom tokens, which is more advantageous.

\subsection{Streaming Training of the User-Order Source Model}
During the streaming training stage, the input to the Source Instance Transformer (SIT) is a sequence where the most recent element is InsEmb of the current instance (generated by the compression layer), and the remaining elements are InsEmb of historical instances belonging to the same user. The sequence is ordered with the current instance's InsEmb as the first element, followed by historical instances in reverse chronological order. SIT strictly adheres to the causal constraint to ensure that each element can only access the InsEmb of past instances. The body text only provides one approach to obtaining the historical InsEmb, while we detail another one and their strength and weakness in the following. 

To obtain historical InsEmb required by SIT, two practical approaches are proposed with distinct trade-offs:
\begin{itemize}[leftmargin=*]
    \item \textbf{Method 1: Store Retrieval}. Directly retrieving the InsEmb of the previous $T-1$ instances of the user from PS (where $T$ is the maximum sequence length). 
    \begin{itemize}[leftmargin=*]
        \item Advantages: Low computational cost and high speed, suitable for streaming training scenarios.
        \item Disadvantages: Relying on historical InsEmb generated by past training iterations, which may inconsistent with the latest model parameters.
    \end{itemize}
    \item \textbf{Method 2: Recomputation}. Retrieving the original features of the previous $T-1$ instances and recompute their InsEmb via forward propagation of the source model.
    \begin{itemize}[leftmargin=*]
        \item Advantages: Using the latest model parameters to recompute InsEmb, and eliminating discrepancies from stale historical values and ensuring high modeling accuracy. Specifically, only forward propagation is allowed for historical InsEmb without gradient backpropagation, while the current InsEmb retains normal gradient flow.
        \item Disadvantages: Significantly increased computational cost (e.g., 256x more computations when $T=256$), only applicable for scenarios with enough computation resources or with a lower requirement of latency.
    \end{itemize}
\end{itemize}
Given that Method 2 introduces excessively high computational overhead due to the recomputation of historical InsEmb, we adopt Method 1 (Store Retrieval) as the default approach in practice.

\begin{figure}[tb]
  \centering
  \includegraphics[width=\linewidth]{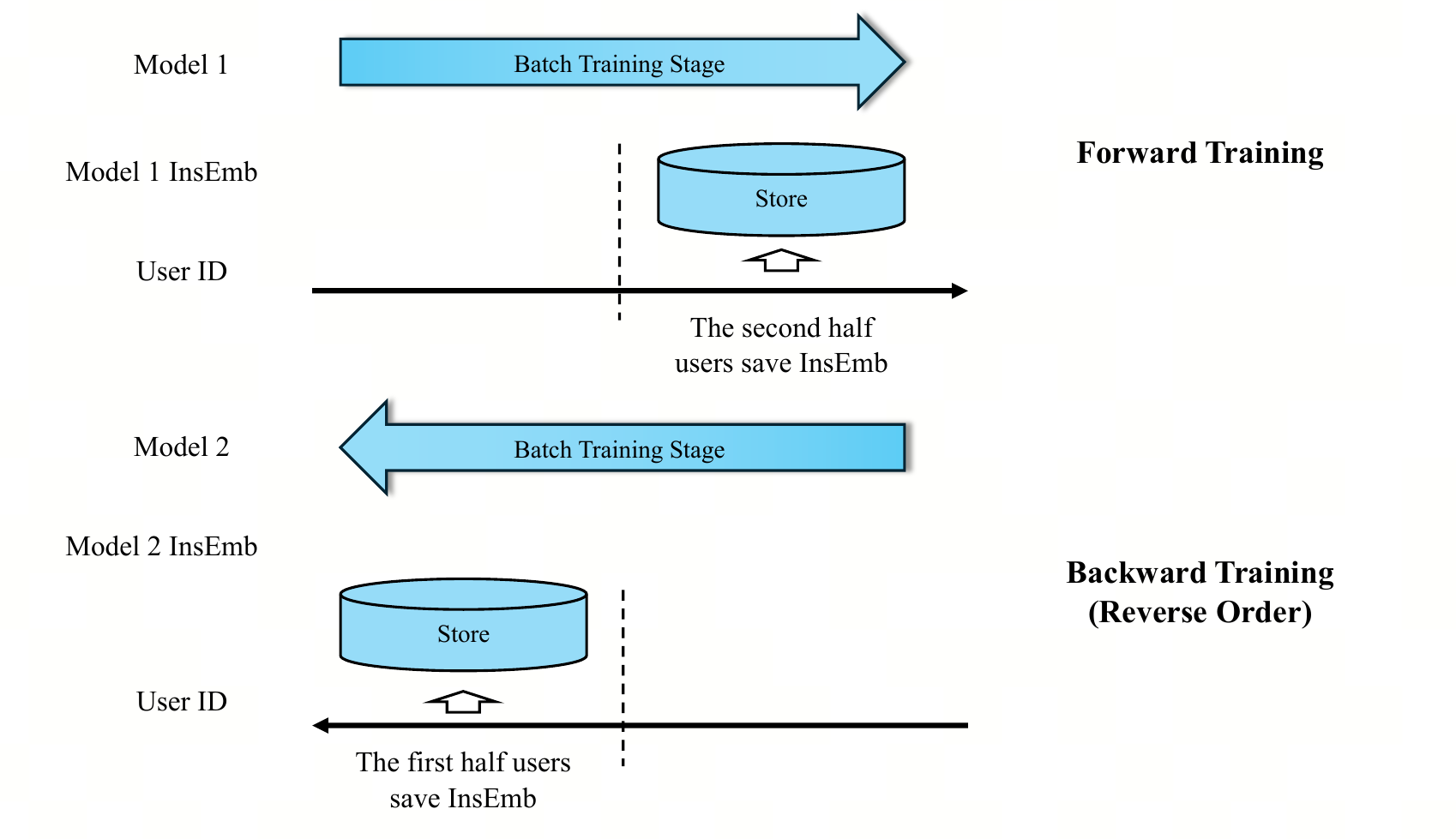}
  \caption{An enhancement training approach for the user-order source model.}
  \label{fig:iat_forward_backward}
\end{figure}

\subsection{Further Performance Enhancement of the User-Order Source Model}
Finally, we consider a more complex training approach to enhance the performance of the user-order source model. Because the training instances are ordered by the user during the batch training stage of the user-order source model, the earliest trained users may produce weak InsEmb because the source model may not converge, while the finally trained users' InsEmb could benefit from a converged and better source model. That is, the one-pass user-order training may lead to unfairness among different users. Hence, we propose to train the user-order source model in two separate jobs, and take the exactly reversed order for these two jobs. The illustration is provided in Fig.~\ref{fig:iat_forward_backward}. One model processes training instances in the normal order but only stores the InsEmb of the later user IDs. In contrast, another model processes training instances in the reversed user order. We also explore some experimental studies to verify the advantage of this two-pass paradigm. The one-pass user-order source model itself obtains about an AUC gain of $0.6\%$, while the proposed two-pass one further obtains a gain up to $0.8\%$.

\end{document}